\documentclass[a4paper]{PoS}

\usepackage{amsmath}
\usepackage{amssymb}

\usepackage{graphicx}
\usepackage{ctable}

\def\apj{ApJ, }%
\def\aap{A\&A, }%
\def\apjs{ApJS, }%
\def\apjl{ApJ, }%
\def\aj{AJ, }%
\def\mnras{MNRAS, }%
\def\pwn{HESS\,J1303$-$631}
\def\pulsar{PSR\,J1301$-$6305}

\def\gammaray{$\gamma$-ray}
\def\gammarays{$\gamma$-rays}

\def\hess{H.E.S.S.}

\def\xmm{\textit{XMM-Newton}}

\def\fermi{\textit{Fermi}-LAT}
\def\atca{\textit{ATCA}}
\def\deg{$^{\circ}$}

\title{Radio Observations of the Pulsar Wind Nebula \pwn\ Field of View with \atca}

\ShortTitle{\atca\ Observations of \pwn}

\author{\speaker{Iurii Sushch}\\
        Centre for Space Research, North-West University, Potchefstroom 2520, South Africa\\
        Astronomical Observatory of Ivan Franko National University of L'viv, vul. Kyryla i Methodia, 8, L'viv 79005, Ukraine\\
        E-mail: \email{iurii.sushch@nwu.ac.za}}

\author{Igor Oya\\
        DESY Zeuthen, Platanenallee 6, D-15738 Zeuthen, Germany}

\author{Ullrich Schwanke\\
Institut f\"{u}r Physik, Humboldt-Universit\"{a}t zu Berlin, Newtonstr. 15,
D 12489 Berlin, Germany}

\author{Simon Johnston\\
Australia Telescope National Facility, CSIRO, PO Box 76, Epping, NSW 1710, Australia}

\author{Matthew L. Dalton\\
Active Space Technologies GmbH, Carl Scheele Strasse 14, 12489 Berlin, Germany}

\abstract{Based on its energy-dependent morphology the initially unidentified very high energy (VHE; $E \gtrsim 100$ GeV) gamma-ray source \pwn\ was recently associated with the pulsar \pulsar. Subsequent detection of X-ray and GeV counterparts also supports the identification of the \hess\ source as evolved pulsar wind nebula (PWN). We report here on recent radio observations of the \pulsar\ field of view (FOV) with ATCA dedicated to search for the radio counterpart of this evolved PWN. Observations at 5.5 GHz and 7.5 GHz do not reveal any extended emission associated with the pulsar. The analysis of the archival 1.384 GHz and 2.368 GHz data also does not show any significant emission. The 1.384 GHz data reveal a hint of an extended shell-like emission in the \pulsar\ FOV which might be a supernova remnant. We discuss the implications of the non-detection at radio wavelengths on the nature and evolution of the PWN as well as the possibility of the SNR candidate being the birth place of \pulsar.}

\FullConference{The 34th International Cosmic Ray Conference,\\
		30 July- 6 August, 2015\\
		The Hague, The Netherlands}

\begin{document}

\section{Introduction}
\pwn\ is one of the most prominent examples of the so-called very
  high energy (VHE; $E>100$\,GeV) \gammaray\ "dark" sources, those
  which were detected in the VHE band but did not have counterparts
at other energy bands. It was discovered in 2005
\cite{2005A&A...439.1013A} but the nature of the source was unclear
until 2012, when a detailed study of the energy-dependent morphology
provided evidence of the association with the pulsar \pulsar\ 
\cite{2012A&A...548A..46H}. With the increase of the energy threshold
a very extended emission region  ($\sim 0.4$\deg$\times 0.3$\deg\ at the 
$(0.84 - 2)$ TeV band)  of VHE \gammarays\ "shrinks" towards the
position of the pulsar at $E> 10$ TeV. Such an energy-dependent 
morphology is expected 
for ancient pulsar wind nebulae (PWNe), which feature several
populations of relativistic electrons generated by the pulsar.  Young
electrons located close to the pulsar are not cooled yet and, thus,
very energetic. These energetic electrons generate
the VHE emission around the pulsar via Inverse Compton (IC)
scattering on the cosmic microwave background photons (CMB). 
Older, cooled down, lower energy electrons might be spread farther 
away from the pulsar for several reasons (e.g. proper motion of the 
pulsar which causes that older particles are left behind and/or particle 
diffusion) but they can still produce \gammarays\ via IC scattering, however 
at lower energies than the young electrons.

The association of \pwn\ with the pulsar is further supported 
by the detection of its X-ray counterpart with \xmm\ 
\cite{2012A&A...548A..46H}. The size of the X-ray PWN is much 
smaller than the size of the VHE source, extending $2^\prime-3^\prime$ from the 
pulsar position towards the center of the VHE \gammaray\ emission region. The
much smaller size of the X-ray emitting region can be explained by an effective 
synchrotron cooling of older electrons to energies too low to generate 
synchrotron emission in the X-ray energy range and/or due to the decreasing of 
the magnetic field strength in the PWN with time 
(see e.g. \cite{2009arXiv0906.2644D}). The tail-like extension of the 
X-ray source might be an indication of the proper motion direction of the 
pulsar triggering speculations about its possible birth-place 
\cite{2012A&A...548A..46H}. 

A careful analysis of archival data from the 
Parkes-MIT-NRAO (PMN) survey at 4.85 GHz 
\cite{1993AJ....106.1095C} revealed also 
a hint of radio emission at the pulsar position with size 
comparable to the X-ray emission region. Data analysis showed a 
$\sim3\sigma$ feature with a peak flux of $0.03$ Jy/beam which is at the 
detection limit of the survey \cite{2012A&A...548A..46H}. This hint of a 
radio counterpart of \pwn\ triggered new dedicated observations with the 
Australian Telescope Compact Array (\atca), which were conducted in 
September 2013. Results of these observations are presented in this paper.

Recently, the counterpart of \pwn\ was finally detected at GeV
energies with \fermi\ \cite{2013ApJ...773...77A}. The source is
contaminated by the emission of the nearby Supernova remnant (SNR)
Kes\,17, but it is clearly seen above $31$\,GeV. The emission region of the 
GeV counterpart of \pwn\ is as expected larger than the TeV source, but
the morphology of the emission region is very similar and features an
extension in the same direction as the TeV source.

\section{Observations and Data Analysis Results}
\subsection{Observations and Data Description}

The \atca\ observations of the field of view (FOV) around \pulsar\ 
were conducted on September 5th, 2013.  Observations were performed with
the $1.5$A configuration of the array at $5.5$ and $7.5$\,GHz
frequencies and centered at $\alpha =
13^{\mathrm{h}}02^{\mathrm{m}}10.0^{\mathrm{s}}$, $\delta =
-63^{\circ}05^{\prime}34.8^{\prime\prime}$ (J2000.0). The total time
of observations was $2626.7$ minutes. Primary and secondary calibrators
were $1934-638$ and $1352-63$, respectively. All the details of the collected 
data are listed in Table \ref{archive_data}.

In this paper we also considered archival \atca\ data obtained during 
observations of \pwn\, centred at $\alpha =
13^{\mathrm{h}}03^{\mathrm{m}}0.40^{\mathrm{s}}$, $\delta =
-63^{\circ}11^{\prime}11.55^{\prime\prime} $, and performed in the $1.384$\,GHz and
$2.368$\,GHz bands.
The archival data used in the analysis, taken as part of the 
Reinfrank\,et\,al. project C1557, are presented in Table \ref{archive_data}. 
Only the archival data taken with all 6 antennas and with observational time 
longer than 100 minutes were used in the analysis.

\begin{table*}
\centering
\caption{Details of the \atca\ data of the \pwn\ FoV analysed in
  this paper}
\label{archive_data}
\begin{tabular}{c c c c c c }
\hline
\hline
\\
Date & Right Ascention& Declination& Time, [min]&Array& Frequencies, [MHz]\\
\hline
\\
2013-Sep-05& $13^\mathrm{h}2^\mathrm{m}10.00^\mathrm{s}$& $-63^{\circ}5^\prime34.80^{\prime\prime}$&2626.7& 1.5A& 5500, 7500\\
2006-Oct-25& $13^\mathrm{h}3^\mathrm{m}0.40^\mathrm{s}$& $-63^{\circ}11^\prime11.55^{\prime\prime}$&433.8&EW352&1384, 2368\\
2007-Mar-13& $13^\mathrm{h}3^\mathrm{m}0.40^\mathrm{s}$& $-63^{\circ}11^\prime11.55^{\prime\prime}$&618.1&750D&1384, 2368\\
2007-Apr-24& $13^\mathrm{h}3^\mathrm{m}0.40^\mathrm{s}$& $-63^{\circ}11^\prime11.55^{\prime\prime}$&651.8&1.5C&1384, 2368\\	
\hline
\end{tabular}
\end{table*}

\subsection{Data Analysis}

\begin{figure}
\centering
\resizebox{\hsize}{!}{\includegraphics{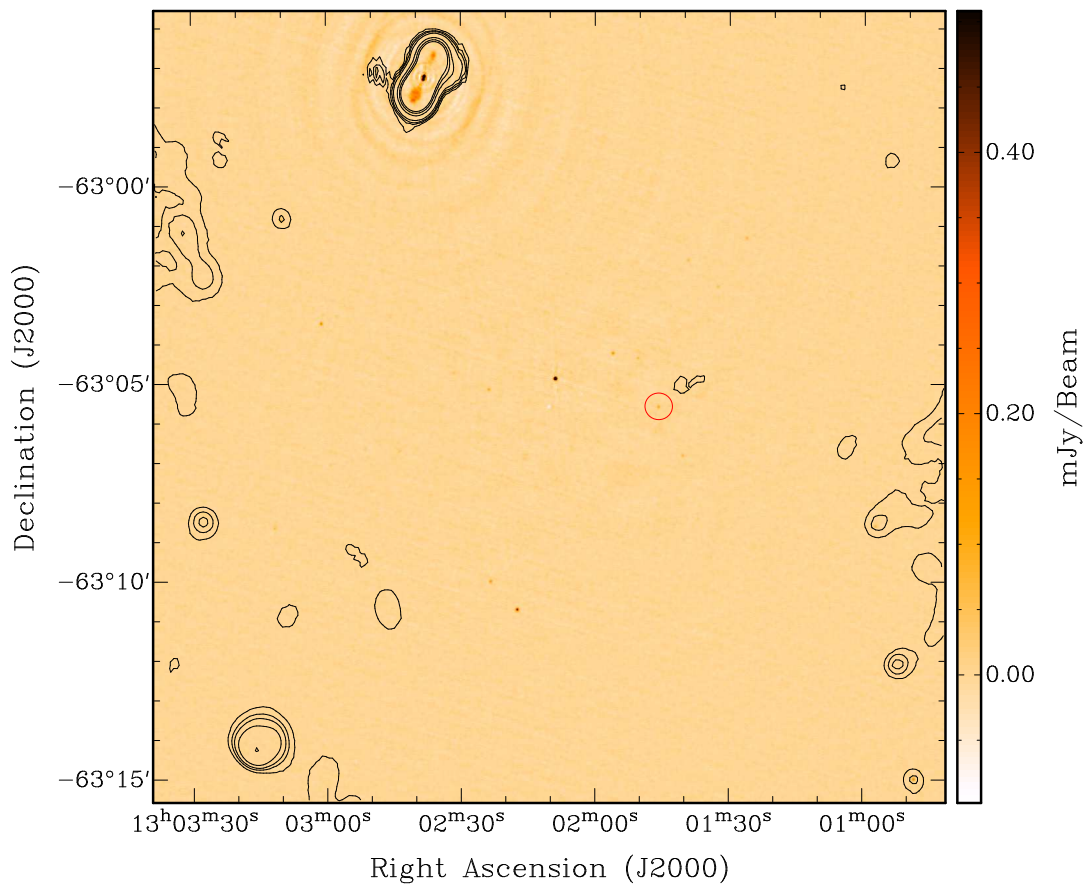}}
\caption{Radio flux map of the \pwn\ FOV at $5.5$\,GHz
  overlaid with significance contours from the  $1.384$ GHz band 
(from $2\sigma$ to $6\sigma$ with the interval of $2\sigma$ and from $10\sigma$ to $70\sigma$ with 
the interval of $20\sigma$). The red circle indicates the position of the pulsar PSR J1301-6305.}
\label{radiomap}
\end{figure}

The data analysis was performed using the \texttt{miriad}
package \cite{1995ASPC...77..433S}. The resulting clean 
radio image at $5.5$\,GHz is shown in Fig.\,\ref{radiomap}. 
The pulsar \pulsar\ is detected at
the position $\alpha = 13^\mathrm{h}01^\mathrm{m}45.680^\mathrm{s} \pm
0.012^\mathrm{s}$, $\delta = -63^{\circ}05^\prime34.848^{\prime\prime}
\pm 0.196^{\prime\prime}$.  No significant extended emission
coincident with the pulsar position was detected. The fitted image
root mean square (RMS) noise is calculated using the \texttt{imsad} task at
the level of $6.072 \times 10^{-6}$\,Jy/beam and the major and minor
axes of the beam are estimated to be $3.793^{\prime\prime}$ and
$3.652^{\prime\prime}$, respectively.

There is no extended emission coincident with the pulsar position
detected at $7.5$\,GHz as well. The fitted image RMS noise,
calculated using the \texttt{imsad} task, is at the level of $6.314
\times 10^{-6}$\,Jy/beam and the major and minor axes of the beam are
estimated to be $3.055^{\prime\prime}$ and $2.900^{\prime\prime}$,
respectively.

\begin{figure}
\centering
\resizebox{\hsize}{!}{\includegraphics{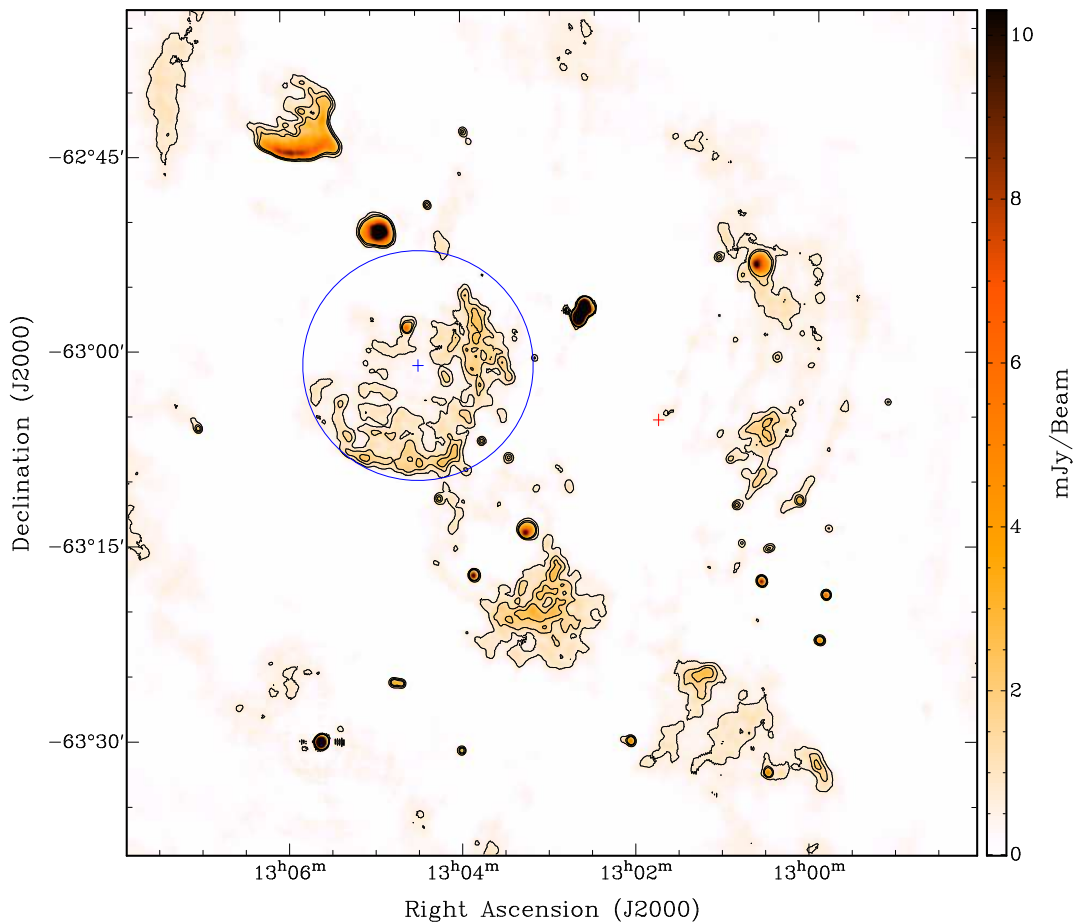}}
\caption{Radio flux map of the \pwn\ FOV at $1.384$\,GHz. Contour levels represent $2\sigma$,$4\sigma$, and $6\sigma$  $1.384$\,GHz emission. The red cross indicates the position of the pulsar PSR J1301-6305. The blue circle shows the result of the disk fit to the SNR candidate and the blue cross indicates the centre of the disk.}
\label{radiomap1384}
\end{figure}

The $1.384$\,GHz (Fig.\,\ref{radiomap1384}) and $2.368$\,GHz flux maps which combine all the archival data 
listed in Table \ref{archive_data} also do not reveal any significant emission 
coincident with the pulsar. The observations at $1.384$\,GHz, however, reveal a shell-like structure to
the east of the pulsar position (Fig. \ref{radiomap1384}) which might be 
an SNR. The SNR FoV was fit with the 
two-component model of the point-source and a circular disk using 
the \texttt{miriad} task \texttt{imfit} (Fig.\,\ref{radiomap1384}). The point-source component was used to 
subtract the emission from a source $\sim3^{\prime}$ to the north 
from the SNR candidate centre (blue cross on Fig. \ref{radiomap1384}). 
This point-like source features a $\sim10$ times higher peak flux than 
anywhere else in the SNR candidate and is, thus, probably a separate source. The disk fit results in a total integrated flux of 
$0.692\pm0.005$\,Jy with a peak value of $ (6.87 \pm  0.17)\times10^{-4}$\,
Jy/beam. The diameter of the disk is estimated to be $1063^{\prime\prime} \pm 4^{\prime\prime}$. The border 
of the disk is shown as a blue circle in Fig.\,\ref{radiomap1384} with its center marked as a blue cross. 
The coordinates of the disk center are 
$\alpha = 13^{\mathrm{h}}04^{\mathrm{m}}29.059^{\mathrm{s}} \pm 0.035^{\mathrm{s}}$ and 
$\delta = -63^{\circ}01^{\prime}22.140^{\prime\prime} \pm 0.926^{\prime\prime}$. The fitted image RMS noise is estimated
to be $3.383\times 10^{-6}$\,Jy/beam and major and minor axes of the
beam are estimated to be $29.33^{\prime\prime}$ and
$26.49^{\prime\prime}$, respectively.

\begin{figure}
\centering
\resizebox{\hsize}{!}{\includegraphics{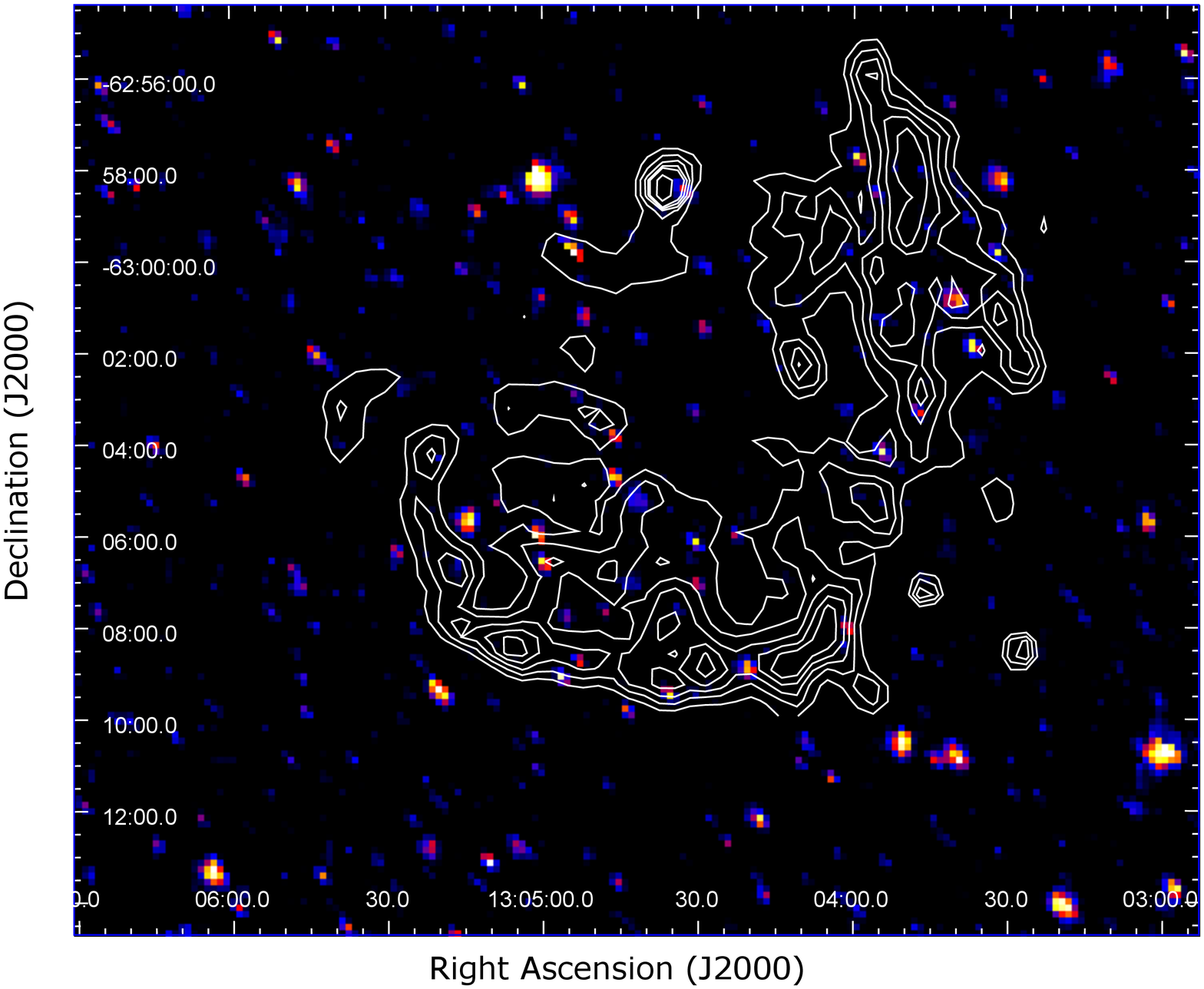}}
\caption{The infrared image of the SNR candidate FoV obtained from the 2MASS survey in the H-band 
($1.65\,\mu$m) \cite{2006AJ....131.1163S} overlaid with $1.384$\,GHz radio contours.}
\label{irsnr}
\end{figure}

Identification of the detected shell-like structure as an SNR is supported by 
infrared and optical observations which do not show any extended thermal emission 
from that region, suggesting that the observed radio emission is non-thermal as 
expected for SNRs. Figure \ref{irsnr} shows an infrared image obtained from 
The Two Micron All-Sky Survey (2MASS) in the H-band  ($1.65\,\mu$m) \cite{2006AJ....131.1163S} 
overlaid with $1.384$\,GHz contours. Other infrared surveys as well as optical surveys at various 
bands do not show anything extra.

\section{Discussion}

\subsection{Implications of the lack of radio emission on the PWN evolution}

The lack of detection of a radio counterpart of the evolved PWN \pwn\ does not allow 
to draw unambiguous conclusions and leaves plenty of space for speculations. The main uncertainty 
is the size of the putative radio PWN which might be larger than the FoV which prevents a detection. 
Below we examine two extreme scenarios which can lead to the non-detection 
of a radio counterpart. 

The size of the GeV emission region \cite{2013ApJ...773...77A} indicates that a high amount 
of relatively low energy relativistic electrons is spread out to large distances from the pulsar. 
The same electrons are also responsible for the generation of the radio emission via the synchrotron mechanism 
(see e.g. \cite{2013ApJ...773..139V} and references therein). The radio spectrum of the PWN is expected to follow 
the same shape as the GeV spectrum in the energy range where it can be well described by a power-law. 
Using only two lower energy data points of the GeV spectrum (see Fig.\,8 in \cite{2013ApJ...773...77A}) one can estimate a 
photon index of $1.01 \pm 0.44_{\mathrm{stat}} \pm 0.71_{\mathrm{syst}}$ which corresponds to an electron 
spectral index of $p = 1.02$ which is well compatible with typical values of $1.0-1.3$ for 
the electron population responsible for the radio emission in PWNe (e.g. \cite{2013ApJ...773..139V, 1978AA....70..419W}).
Therefore, it can be expected that the radio emission region should be at least as big as the GeV emission region, 
i.e. with an extension of $0.9^{\circ}$ \cite{2013ApJ...773...77A}. In this case the putative radio PWN would be 
larger than the $5.5/7.5$\,GHz FoV and comparable to the $1.384/2.368$\,GHz FoV and, thus, the emission from the 
PWN cannot be detected in observations reported in this paper.

On the other hand, it is expected that the magnetic field is amplified in the vicinity of the pulsar being frozen 
into the outflowing wind, while at large distances from the pulsar the magnetic field can be already relaxed 
with strength values comparable to the interstellar medium magnetic field of about $3\mu$G supressing the synchrotron emission. 
In this case the size of the putative radio PWN is defined by the region of higher magnetic field and not by the existence of 
relativistic electrons. Assuming the size of the X-ray PWN is completely defined by the region of the high 
magnetic field one might expect a putative radio PWN to be of the same size. This is, however, an extreme 
case as the size of the X-ray PWN might be also limited by the existence of the very high energy electrons 
which have a short lifetime and are expected to be only in the immediate vicinity of the pulsar. Therefore, 
this size should be treated as a lower limit for the region of the higher magnetic field. Assuming the 
size of the X-ray PWN as reported in \cite{2012A&A...548A..46H} the flux upper limit on the radio flux 
at $5.5$\,GHz was estimated using a bias-corrected bootstrap method 
(see e.g. \cite{bootstrap}) based on the measured counts in the region of the X-ray PWN and performing 20.000 simulations. The upper limit at 
$99\%$ confidence level is estimated to be 0.2 mJy. This upper limit is much more constraining (by two 
orders of magnitude) than the one reported in \cite{2012A&A...548A..46H} for the insignificant radio 
feature of a similar size detected at $4.85$\,GHz. 
Assuming that the same electrons produce radio radiation via the synchrotron mechanism and GeV emission via inverse Compton 
(IC) scattering on the Cosmic Microwave Background (CMB) and assuming that the magnetic field is uniform across the 
putative radio PWN one can estimate an upper limit on the magnetic field. For these calculations the total flux in the GeV range 
was scaled to the region of the size of the X-ray PWN with extension of $\sim176^{\prime\prime}$ \cite{2012A&A...548A..46H} assuming 
that the GeV emission is uniformly distributed. 
This yields an upper limit of $500\mu$G for assumed $p = 1.02$. Similarly one can calculate the magnetic field 
strength comparing X-ray and TeV emission. In this case we also assume that X-ray and gamma-ray emission is produced 
by the same population of electrons the energy distribution of which follows a simple power-law. The total TeV flux 
is scaled to the region size of the X-ray PWN and two different values of the electron spectral index are used 
$p = 3.0$ and $p = 3.88$ basing on the estimates of the photon indices of the X-ray, $\Gamma = 2.0^{+0.6}_{-0.7}$, and TeV, 
$\Gamma = 2.44\pm0.03$, spectra \cite{2012A&A...548A..46H}. The obtained values of the magnetic field 
strength are $10\mu$G and $15\mu$G, respectively. This approach was checked comparing X-ray flux to the 
total TeV flux yielding magnetic field estimates of $1.2\mu$G and $2.8\mu$G, respectively, which are 
compatible with an estimate of $1.4\mu$G obtained in \cite{2012A&A...548A..46H}. However, it should be 
noted that a simple scaling of the total TeV flux is not correct as the TeV emission features a strong 
energy-dependent morphology, suggesting that there are much more high energy electrons in the immediate 
vicinity of the pulsar. This fact would decrease the estimate of the magnetic field strength. Nevertheless, 
these estimates are in agreement with the non-detection of the radio emission for the assumed 
electron spectral index $p = 1.02$. For further investigation of the spectral energy distribution of the 
source deep observations in the TeV band are extremely important as they might give an opportunity 
to quantitively distinguish spectral properties in different regions of the PWN.

\subsection{The SNR candidate - the birth place of \pulsar?}

In case the extended source with a shell-like morphology detected at 1.384 GHz is an SNR it 
could be the birth place of the pulsar \pulsar. Assuming a distance to the pulsar of $6.6$\,kpc 
\cite{2012A&A...548A..46H}, the angular size of the SNR candidate of $1063^{\prime\prime}$  (see above) 
corresponds to 34\,pc in diameter. The Sedov solution \cite{1959sdmm.book.....S}, which describes 
the hydrodynamical expansion of an SNR in the adiabatic stage of evolution into the homogeneous medium, 
provides an estimate of the SNR age for a given size of the remnant
\begin{equation}
t_{\mathrm{age}} = 18 \left(\frac{E}{10^{51}\,[\mathrm{erg}]}\right)^{-2}\left(\frac{n_{\mathrm{ISM}}}{1\,[\mathrm{cm}^3]}\right)^{2}\left(\frac{R}{17\,[\mathrm{pc}]}\right)^{5/2}\,[\mathrm{ky}],
\end{equation}
where $E$ is the explosion energy, $n_{\mathrm{ISM}}$ is the number density of the interstellar 
medium and $R$ is the radius of the remnant. This value is somewhat larger 
than the characteristic age of the pulsar of 11 ky \cite{2005AJ....129.1993M}. 
The characteristic age is defined as an upper limit of the true age of the pulsar under the assumption that the 
braking index is $n=3$ (spin-down of the pulsar via dipole radiation). It is calculated in a limit of initial 
spinning period much lower than the current period, $P_0 \ll P$. However, it is known that the braking index might 
be considerably lower than $3$ (see e.g. \cite{2006MNRAS.372..489A, 2011ApJ...741L..13E}) which increases 
the upper limit for the true age of the pulsar. In this case, the true age of the pulsar can be higher than 
the characteristic age. It should also be noted that the SNR age estimate is strongly dependent on the 
ambient medium density which is often considerably lower than $1\,\mathrm{cm}^3$. 

The angular distance between the pulsar and the SNR candidate of $0.32^{\circ}$ corresponds to a 
projected distance of $36$\,pc assuming the distance to the pulsar of $6.6$\,kpc. This corresponds 
to a lower limit on the pulsar velocity of $V_{\mathrm{p}}\geq 1000$\,km/s for the estimate of the SNR age presented above. The pulsar is 
located outside the shell of the SNR candidate, which means 
that in case the SNR 
candidate is indeed the birth place of \pulsar, the pulsar has already escaped the remnant and continues to 
propagate in the ambient medium. This fact together with the large size of the GeV/TeV PWN (larger than the 
SNR candidate) suggests that there should be some evidence of distortion of the shell caused by the escape 
of the pulsar. However, the shell of the SNR candidate does not show strong evidence of distortion on the side 
of the pulsar. This can be naturally explained if the pulsar is not moving in the projected plane but its 
velocity has a considerable perpendicular to the projected plane component. In this case the distorted part 
of the shell is facing the observer and is thus not visible. This configuration, however, requires a 
higher pulsar velocity depending on the angle $\phi$ between the line of sight and the pulsar proper motion 
direction 
\begin{equation}
V_{\mathrm{p}} = V_{0}/ \sin{\phi},
\end{equation}
where $V_{0} = 1000$\,km/s is the lower limit of the pulsar velocity calculated for $\phi = 90^{\circ}$. In any 
case the association of the pulsar with the SNR candidate would place \pulsar\ among the fastest known pulsars with 
highest detected velocity of $\sim1600$\,km/s \cite{1998ApJ...505..315C}.

\end{document}